\documentclass[twocolumn,nofootinbib,amsmath,amssymb,preprintnumbers]{revtex4-1}

\usepackage{graphicx}
\usepackage{slashbox}
\usepackage{color}
\usepackage{epstopdf}

\newcommand{\squishlist}{
   \begin{list}{$-$}
    { \setlength{\itemsep}{0pt}      \setlength{\parsep}{3pt}
      \setlength{\topsep}{4pt}       \setlength{\partopsep}{0pt}
      \setlength{\leftmargin}{1.5em} \setlength{\labelwidth}{1em}
      \setlength{\labelsep}{0.7em} } }

\newcommand{\squishlisttwo}{
   \begin{list}{-}
    { \setlength{\itemsep}{0pt}    \setlength{\parsep}{0pt}
      \setlength{\topsep}{0pt}     \setlength{\partopsep}{0pt}
      \setlength{\leftmargin}{2em} \setlength{\labelwidth}{1.5em}
      \setlength{\labelsep}{0.6em} } }

\newcommand{\squishend}{
    \end{list}  }

\newcommand{\vol}[1]{\textbf{#1}}

\newcommand{\tarttitle}[1]{``#1'',} 

\newcommand{\tref}[1]{(\ref{#1})}

\newcommand{\bea}{\begin{eqnarray}}
\newcommand{\eea}{\end{eqnarray}}
\newcommand{\beq}{\begin{equation}}
\newcommand{\eeq}{\end{equation}}

\newcommand{\Tsteady}{\tau} 

\begin{document}
\title{Turnover Rate of Popularity Charts in Neutral Models}
\author{T.S.\ Evans$^{1,2}$}
\author{A.\ Giometto$^{1}$\email{andrea.giometto10@imperial.ac.uk}}
\affiliation{$^1$ Institute for Mathematical Sciences, Imperial College London, London, SW7 2PG, UK \\
$^2$ Theoretical Physics, Imperial College London, SW7 2AZ, UK}

\begin{abstract}
It has been shown recently that in many different cultural phenomena the turnover rate on the most popular artefacts in a population exhibit some regularities. A very simple expression for this turnover rate has been proposed by Bentley et al. \cite{BLHH07} and its validity in two simple models for copying and innovation is investigated in this paper. It is found that Bentley's formula is an approximation of the real behaviour of the turnover rate in the Wright-Fisher model, while it is not valid in the Moran model.
\end{abstract}

\maketitle

\section{Introduction}

In genetics, neutral models in which the different variations provide no intrinsic advantage play a central role. Classic examples are the Wright-Fisher and Moran models \cite{W61,E04} which are Markov processes.  It is not surprising that they are also particular limits of other statistical physics models such as Urn models and zero range processes \cite{GL02,EH05} or network rewiring models \cite{Evans07,EP07} (see \cite{EP07} for further examples of these models).  One of the most interesting applications is to cultural transmission \cite{Neiman95,BM99,HB03,BHS04,HBH04}. In this case the popularity of `artefacts' with no intrinsic value (uniform fitness) varies because individuals change their choice of artefact by either copying the artefact chosen by another individual  (inheritance) or by innovating by choosing a new artefact (mutation). Under this hypothesis a certain cultural trait becomes more popular than others simply through imitation and not because of an intrinsic benefit it provides.  Despite its simplicity, these models can reproduce some of the features of real data sets, such as Neolithic pottery \cite{Neiman95,BHS04}, popular music charts  \cite{BM99}, baby names \cite{HB03,BHS04}, patents \cite{BHS04} and dog breeds \cite{HBH04}.

Much is known about these neutral models, including many exact results \cite{W61,E04}.  However the context of cultural transmission throws up new and unanswered questions since they are of little practical use in other applications. In this paper we study popularity charts, the list of the $y$ most popular artefacts at any one time, and ask how many artefacts enter or leave this list each time the chart is updated, the turnover rate $z$. This is motivated by the work of Bentley et al.\ \cite{BLHH07} who find that in the Wright-Fisher model the turnover rate $z$ of the top $y$ chart is $z=\sqrt{\mu} \cdot y$ where $\mu$ is the innovation rate. This result is interesting because the turnover rate $z$ is independent of the population size $N$ and the square root dependence on $\mu$ is reminiscent of a random walk process that might be addressed with a theoretical analysis.  It is of practical use as sometimes we only have access to data on the most popular artefacts and we may not have a useful sample of the whole population.  In such situations it can be used to provide estimates of the model parameters from a data set \cite{BOB11}.
The aim of this paper is to perform a comprehensive study of the turnover rate in the Wright-Fisher and Moran models.

\section{The Wright-Fisher model}

The model investigated here and in \cite{BLHH07} is one of the family of Wright-Fisher models and is illustrated in Fig. \ref{fWF}.  In it each of $N$ individuals is characterised by an artefact of no intrinsic value (e.g.\ a brand of shoes, a dog breed, a name, etc.). At each time step all individuals in the population are simultaneously assigned a new artefact.  With probability $(1-\mu)$ an individual will copy the artefact choice from the previous time step of an individual selected uniformly at random.  Otherwise with probability $\mu$ an individual innovates by choosing a new artefact.
\begin{figure}
\begin{center}
\begin{tabular}{ccccc}
Individual	& 	&	t	&		& 	t+1	\\ \hline
1	&	&	A	&	$\xrightarrow{\textit{copy 3}}$	&	B	\\
2	&	&	A	&	$\xrightarrow{\textit{copy 2}}$	&	A	\\
3	&	&	B	&	$\xrightarrow{\textit{copy 1}}$	&	A	\\
4	&	&	A	&	$\xrightarrow{\textit{innovate}}$	&	D	\\
5	&	&	C	&	$\xrightarrow{\textit{copy 6}}$	&	A	\\
6	&	&	A	&	$\xrightarrow{\textit{copy 6}}$	&	A	\\
\end{tabular}\\
\vspace{0.2cm}
\end{center}
\textbf{Top 3 chart}:\\
\begin{center}
\begin{tabular}{ccccc}
Position	& 	&	t	&	& 	t+1	\\ \hline
1$^{\textit{st}}$		&	&	A	&		&	A	\\
2$^{\textit{nd}}$	&	&	B	&	$\xrightarrow{\textit{ z = 2}}$	&	B	\\
3$^{\textit{rd}}$		&	&	C	&		&	D	\\
\end{tabular}
\end{center}
\label{fWF}
\caption{A simple representation of the Wright-Fisher model. The artefacts are labelled by letters. In this example two successive time steps are shown for a population of six individuals and the top three chart with a turnover of two ($N=6$, $y=3$, $z=2$).}
\end{figure}

The analytical solutions of this model \cite{E04} show that the frequency of artefacts in a population is typically a power law with a cutoff (at least for $N \gg \mu N \gtrsim 1$) and this has been fitted to data on the frequency of various modern cultural variants \cite{HB03,BHS04,HBH04}.

\subsection*{Definition of turnover}

Our definition of the turnover $z$ in the top $y$ chart, the list of the $y$ most popular artefacts, is defined as the sum of the number of artefacts exiting the top chart plus the numbers of new artefacts entering the top chart at the same time step.
This definition of turnover is slightly different from \cite{BLHH07} where it is defined as the number of new artefacts that enter the top $y$ chart relative to the previous time step. In most situations the difference between the two definitions is given by a factor of $2$.  Our definition is more informative for situations where a artefact exits the top chart by becoming extinct with no new artefacts entering it: in this configuration we have a turnover $z=1$, while the definition in \cite{BLHH07} would have $z=0$.  In our notation the result of \cite{BLHH07} is that
\begin{equation}
z=2 \cdot \sqrt \mu \cdot y
\label{newturnover}
\end{equation}
Bentley et al.\ \cite{BLHH07} find this through numerical analysis of this Wright-Fisher model and also find support for this form in data for baby names and dog breeds.

\subsection*{Simulations of the model}

During one simulation the model starts with every individual assigned to a unique artefact and is then first updated $\tau$ times. After reaching a steady state the frequency of every artefact in the population is computed at each time step and the top $y$ chart is built using the \textit{quicksort} algorithm for the next $T$ steps. The temporal average $\bar z$ of the turnover rate $z$ is then computed by comparing two successive top $y$ charts and recorded. To perform an ensemble average the model is rerun $E$ times\footnote{$E$ is chosen so that the error on $z$ is not larger than 10\%. To reduce computational times the model is run for $\Tsteady$ time steps to reach a steady state only once. Successive iterations start from the last configuration of the previous ones. This procedure doesn't affect the results because the system is in a steady state.} and the ensemble average $\langle \bar z \rangle$ of $\bar z$ is computed. This is the estimate of the turnover rate $z$ that is stored for further analysis along with an estimate of the standard deviation in this measurement.

We started our simulations from a configuration where all the individuals had a different artefact. We checked that our simulations had reached a steady state by studying
\begin{equation}
F_2(t)=\frac{\langle k(k-1) \rangle}{N(N-1)} (t) = \frac{\sum_{k=0}^N k(k-1) \langle n(k,t) \rangle }{N(N-1)}
\label{F2def}
\end{equation}
where $n(k,t)$ is the number of artefacts chosen by $k$ individuals at time $t$ and the symbols $\langle \dots \rangle$ indicate an ensemble average. This quantity can be shown analytically to evolve in time as
\begin{equation}
F_2(t)= F_2(\infty) + \left[ F_2(0)-F_2(\infty) \right] \cdot (\lambda_2)^t
\label{F2evolve}
\end{equation}
with $\lambda_2 = (1-\mu)^2(N-1)/N$ \cite{E04} so that $\tau^{-1} \sim \ln(\lambda_2) \lesssim \textrm{min}((2\mu)^{-1},N)$ (a similar results hold for all eigenvalues).   Given the values used in our simulations we chose $\Tsteady=4 \mu^{-1}$ and ran for $T=50+\mu^{-1}$ time steps ($50$ time steps were added to ensure a minimum amount of time steps even for big values of $\mu$).
\subsection*{Analysis}

Motivated by \cite{BLHH07} we fitted our data for $z$ as a function of $\mu$, $y$ and $N$ to the following form
\begin{equation}
z=d\cdot \mu^a y^b N^c
\label{funcform}
\end{equation}
We looked at around 6000 different parameter values taken from the ranges $\mu \in [5\cdot 10^{-5} , 0.115]$, $y \in [2 , 1411]$ and $N \in [180 , 3993]$ with the constraint that $y<N$. This largely extends the range of values studied in \cite{BLHH07} which come from $\mu \in [2\cdot 10^{-4} , 0.02]$, $y \in [5 , 50]$ and $N \in [500, 4000]$. Estimates for coefficients $a$, $b$, $c$ and $d$ in equation (\ref{funcform}) were obtained using a linear fit to the data for $\ln(z)$.

\begin{figure}[!hbt]
\begin{center}
\includegraphics[height=\columnwidth,angle=270]{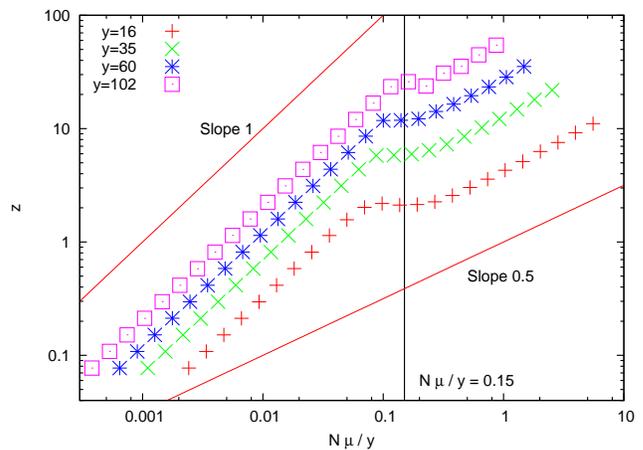}
\caption{Existence of a critical point $N \mu \simeq 0.15 \cdot y$. Plotted curves are for fixed values of $y$ and $N$. Error bars are smaller than symbols.}
\label{critical}
\end{center}
\end{figure}

It is found that the turnover rate $z$ exhibits two different behaviours in the two regions: $N \mu < 0.15 \cdot y$ and $N \mu > 0.15 \cdot y$, as can be seen in figure \ref{critical}, with the transition between these two behaviours occurring around $N \mu \simeq 0.15 \cdot y$.
\begin{figure}[!hbt]
\begin{center}
\includegraphics[height=\columnwidth,angle=270]{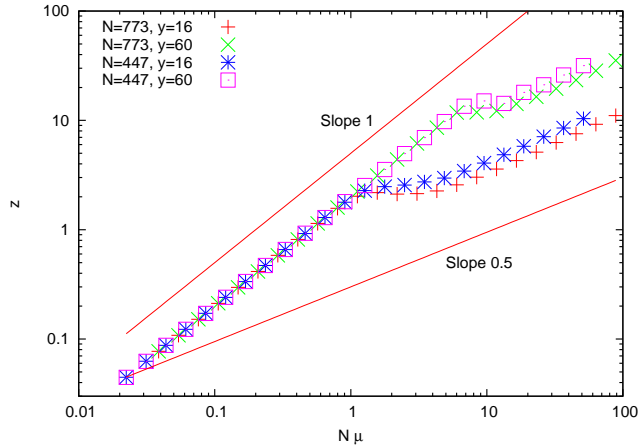}
\caption{Existence of a critical point $N \mu \simeq 0.15 \cdot y$. Plotted curves are for fixed values of $y$ and $N$. In the region $N\mu < 0.15 \cdot y$ all curves collapse in $z=2N\mu$. Error bars are smaller than symbols.}
\label{criticalNmu}
\end{center}
\end{figure}

\subsubsection*{Region $N\mu < 0.15 \cdot y$:}
The observed behaviour of the turnover rate $z$ in this region is: $z\propto \mu$ (see figure \ref{critical} and figure \ref{criticalNmu}). Fitting the data points in this region with the functional form (\ref{funcform}) the following set of values for the fitting parameters is obtained:
\beq
\begin{array}{ll}
a=0.99999(4)  , &
b=-0.00004(7) , \\
c=1.0003(2) , &
d=1.997(2) ,
\end{array}
\label{WFreslowmuN}
\eeq
that is to say:
\begin{equation}
z=2 \cdot N\mu
\end{equation}
within two standard deviations. Note that in this region $z$ is independent on the top chart size $y$. This is best seen in figure \ref{criticalNmu}, where the turnover rate is plotted against the product $N \mu$.

\begin{figure}[!hbt]
\begin{center}
\includegraphics[height=\columnwidth,angle=270]{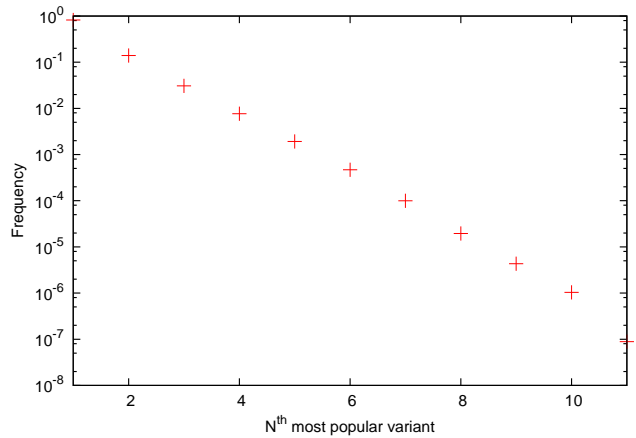}
\caption{Average frequency of artefacts (from the most popular one to the least popular). In the steady state there are no more than 11 artefacts in the population. With $N=180$, $\mu=0.001$ and $y=20$ we have $N \mu \ll 0.15 \cdot y$ and the number of artefacts in the population (11) is smaller than the top chart size $y$ (20).}
\label{frequency}
\end{center}
\end{figure}

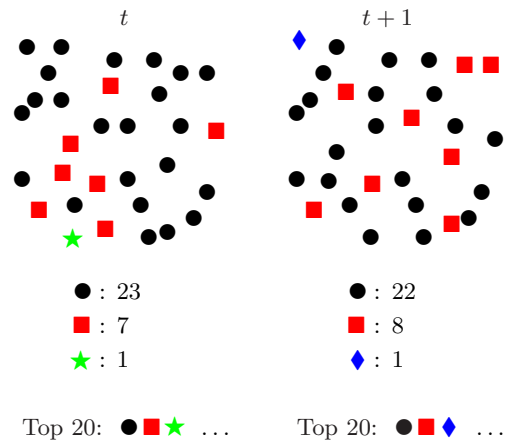
\begin{figure}[!htb]
\begin{center}
\begin{tabular}{ll}
$\ \ \ \ \ \ \ \ \ \ \ \ t$  & $\ \ \ \ \ \ \ \ t+1$\\
\begin{picture}(100, 75)
\color{black}
 \put(20,10){\circle*{6}}
 \put(40,20){\circle*{6}}
 \put(0,20){\circle*{6}}
 \put(30,40){\circle*{6}}
 \put(60,60){\circle*{6}}
 \put(70,15){\circle*{6}}
 \put(5,50){\circle*{6}}
 \put(15,70){\circle*{6}}
 \put(45,10){\circle*{6}}
 \put(35,65){\circle*{6}}
 \put(48,-2){\circle*{6}}
 \put(60,40){\circle*{6}}
 \put(65,5){\circle*{6}}
 \put(52,52){\circle*{6}}
 \put(10,60){\circle*{6}}
 \put(50,65){\circle*{6}}
 \put(0,45){\circle*{6}}
 \put(70,60){\circle*{6}}
 \put(55,25){\circle*{6}}
 \put(15,50){\circle*{6}}
 \put(40,40){\circle*{6}}
 \put(55,0){\circle*{6}}
 \put(2,70){\circle*{6}}
 \color{red}
 \put(25,15){$\blacksquare$}
 \put(3,5){$\blacksquare$}
 \put(30,52){$\blacksquare$}
 \put(12,19){$\blacksquare$}
 \put(28,-2){$\blacksquare$}
 \put(70,35){$\blacksquare$}
 \put(15,30){$\blacksquare$}
 \color{green}
 \put(15,-5){$\bigstar$}
 \end{picture}

 &

 \begin{picture}(10, 75)
\color{black}
 \put(20,10){\circle*{6}}
 \put(40,20){\circle*{6}}
 \put(0,20){\circle*{6}}
 \put(30,40){\circle*{6}}
 \put(70,15){\circle*{6}}
 \put(5,50){\circle*{6}}
 \put(15,70){\circle*{6}}
 \put(45,10){\circle*{6}}
 \put(35,65){\circle*{6}}
 \put(48,-2){\circle*{6}}
 \put(60,40){\circle*{6}}
 \put(65,5){\circle*{6}}
 \put(52,52){\circle*{6}}
 \put(10,60){\circle*{6}}
  \put(30,52){\circle*{6}}
 \put(12,19){\circle*{6}}
 \put(28,-2){\circle*{6}}
 \put(75,35){\circle*{6}}
 \put(15,30){\circle*{6}}
 \put(0,45){\circle*{6}}
 \put(50,65){\circle*{6}}
 \color{red}
 \put(60,60){$\blacksquare$}
 \put(70,60){$\blacksquare$}
 \put(55,25){$\blacksquare$}
 \put(15,50){$\blacksquare$}
 \put(40,40){$\blacksquare$}
 \put(55,0){$\blacksquare$}
  \put(25,15){$\blacksquare$}
 \put(3,5){$\blacksquare$}
 \color{blue}
 \put(-2,70){$\blacklozenge$}
 \end{picture}

\\
&
\\
\hspace{0.7cm}
\color{black}
 \put(0,3){\circle*{6}} \ \ :  23
 &
 \hspace{0.7cm}
\color{black}
 \put(0,3){\circle*{6}} \ \ :  22
 \\
 \hspace{0.7cm}
\color{red}
 \put(-4,0){$\blacksquare$} \ \ \color{black}:  7
 &
 \hspace{0.7cm}
\color{red}
 \put(-4,0){$\blacksquare$} \ \ \color{black} :  8
 \\
 \hspace{0.7cm}
\color{green}
 \put(-5,0){$\bigstar$} \ \  \color{black}:  1
 &
 \hspace{0.7cm}
\color{blue}
 \put(-3,0){$\blacklozenge$} \ \  \color{black}:  1
 \\
 &
 \\
 Top 20:  \put(10,3){\color{black}\circle*{6}} \color{red}\put(15,0){$\blacksquare$} \color{green}\put(23,1){$\bigstar$} \color{black} \hspace{1.2cm} $\dots$
 &
  Top 20:  \put(10,3){\circle*{6}} \color{red}\put(15,0){$\blacksquare$} \color{blue}\put(24,0.5){$\blacklozenge$} \color{black} \hspace{1.2cm} $\dots$
\end{tabular}
\end{center}
\caption{A population of different symbols, the shape being the artefact. With $N\mu \ll y$ only a few artefacts survive in the population and the top chart has empty spots. New artefacts introduced through innovation (on average $N\mu=1$ per generation) enter the top $y$ chart but are extinguished in one time step, producing a turnover equal to $2\cdot N \mu$.}
\label{mechanism}
\end{figure}

This behaviour of the turnover rate $z$ can be explained in the following way: for $N \mu \ll y$ the average number of new artefacts that enter the population in one time step is lower than the top chart size $y$. In this configuration a very small number of artefacts survives in the population in the steady state. It is then likely that the total number of artefacts at a certain time $t$ is lower than the top chart size $y$, as it is confirmed by observations (see figure \ref{frequency}).
In one time step, then, the $N \mu$ new artefacts (on average) introduced in the previous time step (that are in the top $y$ chart) are extinguished through copying, while on average $N \mu$ new artefacts enter the population and the top chart through innovation: the turnover rate $z$ is then equal to $2\cdot N \mu$.
This mechanism is illustrated in figure \ref{mechanism}.

\subsubsection*{Region $N \mu > 0.15 \cdot y$:}

\begin{figure}[!htb]
\begin{center}
\includegraphics[height=\columnwidth,angle=270]{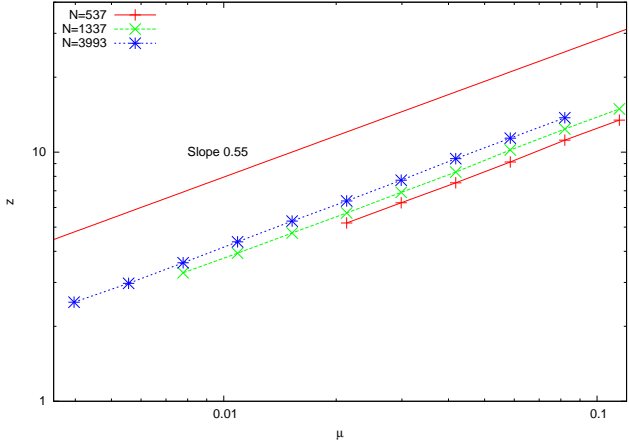}
\caption{Turnover rate z vs $\mu$ for the Wright-Fisher model in the region $N \mu > 0.15 \cdot y$. There is a clear dependence of $z$ on the population size $N$. Error bars are smaller than symbols.}
\label{turnoverz}
\end{center}
\end{figure}

This is the region studied in \cite{BLHH07} and we also find that the dependence of the turnover rate $z$ on the innovation rate $\mu$ is very roughly $z \propto \mu^{\frac12}$, as can be seen in figure \ref{critical}.  However we have also fitted the data in this region to the same functional form $z=d\cdot \mu^a y^b N^c$ \tref{funcform}, using a linear fit to the logarithm of our parameters and $z$.
The resulting values for the fitting parameters $a$, $b$, $c$ and $d$ are:
\beq
\begin{array}{ll}
a=0.550(2) , &
b=0.860(1) , \\
c=0.130(2) , &
d=1.38(2) .
\end{array}
\label{WFreshighmuN}
\eeq
These values are not statistically compatible with the proposed form in equation (\ref{newturnover}) for which $a=1/2$, $b=1$, $c=0$ and $d=2$. The dependence on $\mu$ and $y$ is not so far off that proposed in \cite{BLHH07} ($a$ and $b$ are 10\% and -14\% off the values in \tref{newturnover}) so the practical difference in studying a real data set may be minimal.  However we find a significant dependence on $N$, even with this small power of $c$ we have a 50\% variation in $z$ over the range $N \in [180 , 3993]$.  This dependence on the population size $N$ is clearly seen in figure \ref{turnoverz}.

\begin{figure}[!htb]
\begin{center}
\includegraphics[height=\columnwidth,angle=270]{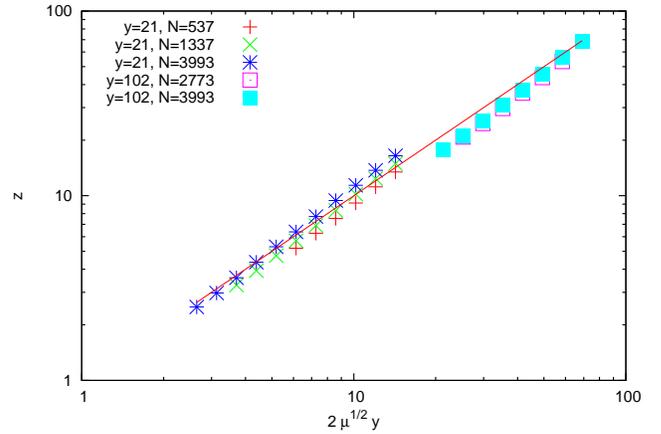}
\caption{Turnover rate $z$ vs $ 2 \sqrt{\mu} y$ for the Wright-Fisher model in the region $N \mu > 0.15 \cdot y$. Error bars are smaller than symbols.}
\label{collapsebentley}
\end{center}
\end{figure}

As a final illustration we plot the turnover rate $z$ against the form (\ref{funcform}) using our best fit values \tref{WFreshighmuN} to create a data collapse. As it can be seen in the figure, collected data lay on the diagonal $z=\mu^a y^b N^c$. The same plot using the form suggested in \cite{BLHH07} is presented in figure \ref{collapsebentley}.
\begin{figure}[!htb]
\begin{center}
\includegraphics[height=\columnwidth,angle=270]{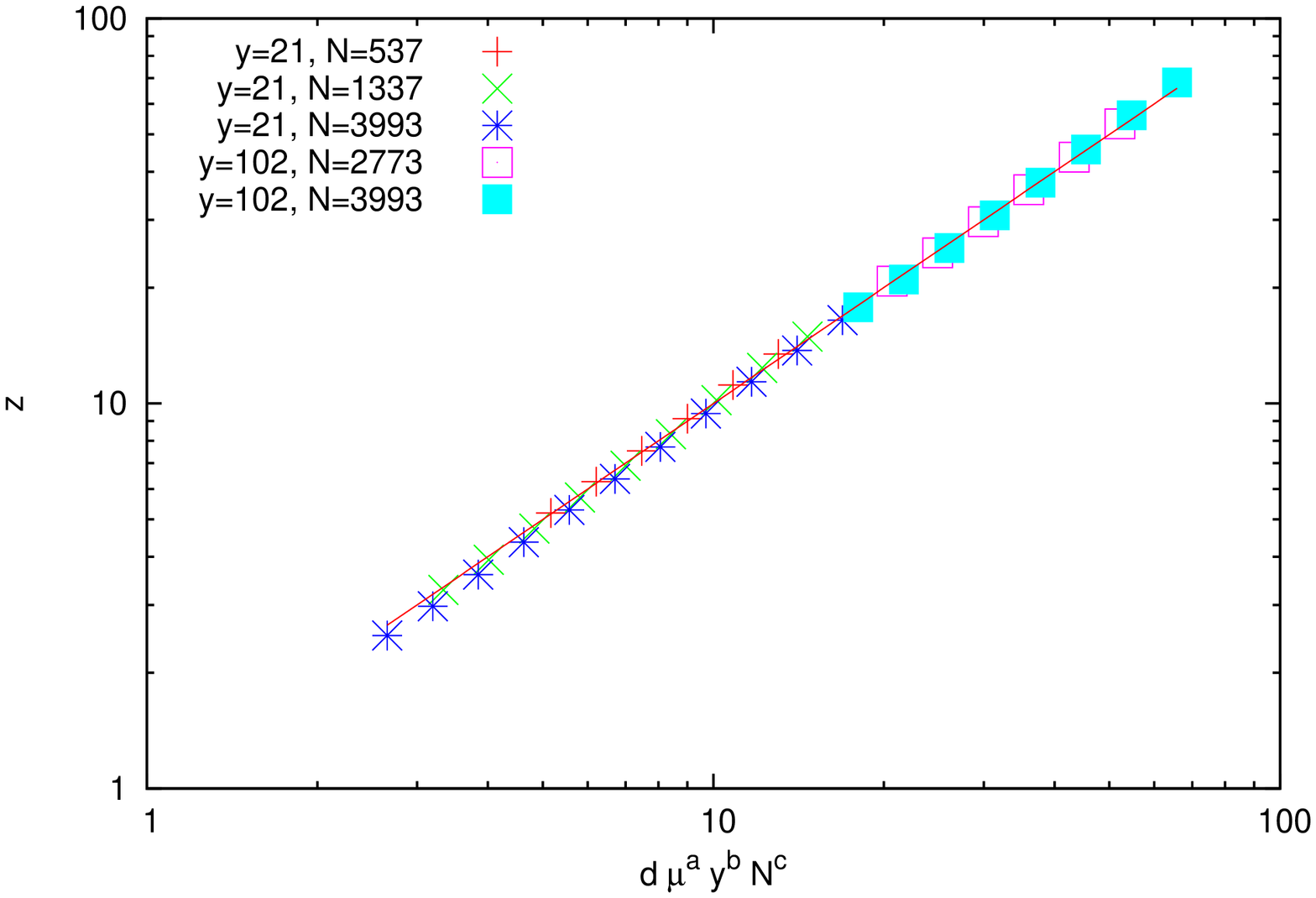}
\caption{Turnover rate $z$ vs $ d \cdot \mu^a y^b N^c$ for the Wright-Fisher model in the region $N \mu > 0.15 \cdot y$. Values used were $a=0.550$, $b=0.860$, $c=0.130$ and $d=1.38$, the best fit values found \tref{WFreshighmuN}. Error bars are smaller than symbols.}
\label{turnovercollapse}
\end{center}
\end{figure}

\subsection*{Residuals}

\begin{figure}[!htb]
\begin{center}
\includegraphics[height=\columnwidth,angle=270]{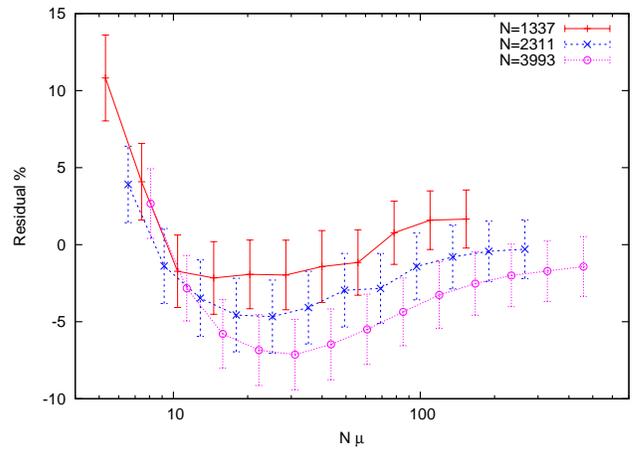}
\caption{Residuals from equation (\ref{funcform}) vs $N \mu$. There is a systematic deviation from the proposed functional form. The deviations for small $N \mu$ are related to data near the critical region $N \mu \simeq 0.15 \cdot y$.}
\label{residuals}
\end{center}
\end{figure}

Residuals from the fit are shown in figure \ref{residuals}. While the peaks for low value of the product $N\mu$ are related to data near the critical point $N\mu \simeq 0.15 \cdot y$, there is an evident systematic deviation from the proposed functional form (\ref{funcform}) for $N \mu \gg 0.15 \cdot y$.

The minimum point of the residuals curves lies in $N \mu \simeq y$ for all values of $N$ and $y$ and increases in absolute value for increasing population size $N$, causing a systematic drift from equation (\ref{funcform}).

We tried many different functional forms and the best fits we found were obtained with the following:
\begin{align*}
&\textrm{Res}(\mu,y,N)=R_0 \cdot \left[1-e^{-f\cdot(\mu N - y)}\right]^2+R_\textrm{min} \\
&R_\textrm{min}(y,N) = A\cdot y + B \cdot N
\end{align*}
where $R_\textrm{min}$ is the value of the residual at the minimum.
The best fit values for the parameters are $R_0=0.062(4)$, $f=0.0096(4)$, $A=0.00034(5)$ and $B=-0.0000193(8)$.  Combining this expression for the residuals with the original form (\ref{funcform}) the following function might be used to describe the behaviour of the turnover rate $z$ of the Wright-Fisher model in the region $N \mu > 0.15 \cdot y$:
\begin{equation}
z = d \cdot \mu^a y^b N^c \cdot \left[ 1 + \textrm{Res}(\mu,y,N) \right]
\label{fitform}
\end{equation}
\begin{figure}[!hbt]
\begin{center}
\includegraphics[height=\columnwidth,angle=270]{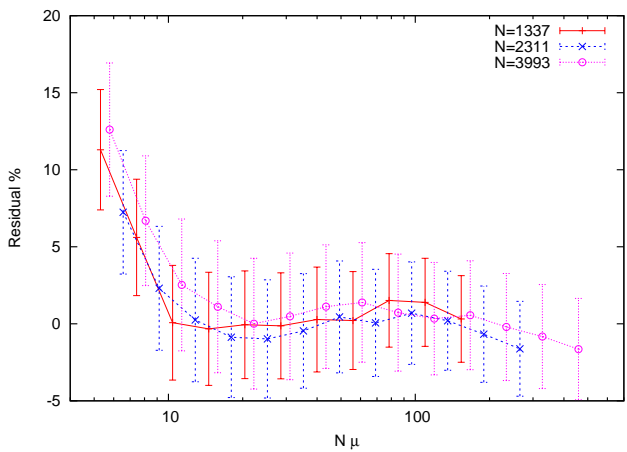}
\caption{Residuals from equation (\ref{fitform}) vs $N \mu$. No systematic deviation from the proposed functional form (\ref{fitform})  can be identified. The deviations for small $N \mu$ are related to data near the critical region $N \mu \simeq 0.15 \cdot y$.}
\label{residualsafter}
\end{center}
\end{figure}
Residuals from equation (\ref{fitform}) are plotted in figure \ref{residualsafter}. The functional form (\ref{fitform}) can now reproduce data within $2\%$ for $N\mu \gg 0.15 \cdot y$ and there is no systematic drift in $N \mu \simeq y$ for increasing values of $N$.

However this form \tref{fitform} is not very satisfactory because of the high number of parameters used to fit the data and because the range of applicability of equation (\ref{fitform}) is strongly limited, due to the negative value of $B$.

\subsection*{Large populations}

The numerical studies of the Wright-Fisher model in \cite{BLHH07} have only a small number of individuals $N \leq 4000$, whereas the real data sets are often much larger e.g.\ around $10^6$ births per year in \cite{HB03}, $6 \times 10^5$ dog breed registrations per year in \cite{HBH04}, the number of albums released in the US (data used in \cite{BM99,BLHH07}) is hard to estimate but the Recording Industry Association of America have suggested that major labels release around 7,000 new CDs every year.

Unfortunately all data sets on which this model has been applied considered population of millions of individuals and this would take too long to simulate. The biggest population size that we have been able to simulate is of the order of 100000 individuals. Results are listed in table \ref{bigdata}.
\begin{table}[htdp]
\begin{center}
\begin{tabular}{c|c|c|c||c|c|c}	
\textbf{$\mu$}	&	\textbf{$y$}	&	\textbf{$N$}	&	$z$ & $z/(2y\sqrt{\mu})$ & $z/z_\textrm{small}$ & $z/z_\textrm{large}$ \\	 \hline
0.0012	&	200	&	100000	&	11.9(1)	& 1.16 & 1.22 & 1.06 \\
0.0024	&	200	&	100000	&	17.6(1)	& 1.11 & 1.21 & 1.05 \\
0.0012	&	400	&	100000	&	24.3(1)	& 1.14 & 1.09 & 0.95 \\
0.0024	&	400	&	100000	&	33.3(2)	& 1.18 & 1.16 & 1.03 \\	\hline
0.0012	&	200	&	120000	&	12.1(1)	& 1.15 & 1.23 & 1.06 \\
0.0024	&	200	&	120000	&	18.0(1)	& 1.09 & 1.21 & 1.05 \\
0.0012	&	400	&	120000	&	23.8(1)	& 1.16 & 1.13 & 0.99 \\
0.0024	&	400	&	120000	&	33.6(2)	& 1.17 & 1.18 & 1.03 \\	\hline
0.0012	&	200	&	144000	&	12.2(1)	& 1.14 & 1.25 & 1.07 \\
0.0024	&	200	&	144000	&	18.3(1)	& 1.07 & 1.22 & 1.05 \\
0.0012	&	400	&	144000	&	23.6(1)	& 1.17 & 1.17 & 1.02 \\
0.0024	&	400	&	144000	&	34.0(2)	& 1.15 & 1.19 & 1.04 \\
\end{tabular}
\end{center}
\caption{Simulation results for big population size $N$.  The last three columns give the ratios of the measured value of the turnover $z$ divided by one of the fitted forms $z = d y^a \mu^b N^c$, respectively $z=2y\sqrt{\mu}$ of \cite{BLHH07} and \tref{newturnover}, $z_\textrm{small}$ best fit for small $N$ of \tref{WFreshighmuN}, and finally $z_\textrm{large}$ best fit for large $N$ of \tref{WFreslargeN}.  All results and best fits are for the $\mu N \gtrsim 0.15 y$ region.}
\label{bigdata}
\end{table}

A surprising result is that for the data in table \ref{bigdata} the simple form suggested in \cite{BLHH07}, $z=2 \sqrt \mu y$, is a better estimate for the turnover rate (typically a 15\% overestimate) than our best fit to $z = d y^a \mu^b N^c$ of (\ref{funcform}) using the data for small $N$ but large $\mu N$ (typically around 20\% overestimate). A factor that might produce this unexpected result is that in all real situations where the model has been tested and in the data in table \ref{bigdata} the size of the top chart is much smaller than the population size: $y \ll N$. We have therefore performed a new data collection in this region (and above the critical point $N \mu \simeq 0.15 \cdot y$).

Approximately 1200 data points have been collected in the range $\mu \in [0.001, 0.13]$, $y \in [5 , 56]$, $N \in [1000, 13000]$
subject to the constraints that $N \mu > y$ (to avoid data from the critical region $N \mu \simeq 0.15 \cdot y$)
and  $y < N / 100$.
The resulting values for the fit $z = d \cdot \mu^a y^b N^c$ \tref{fitform}
are
\beq
\begin{array}{ll}
a=0.558(1) , &
b=0.879(1) , \\
c=0.091(1) , &
d=1.79(2)  .
\end{array}
\label{WFreslargeN}
\eeq
We plot the turnover rate z against the form (\ref{funcform}) using our best fit values (\ref{WFreslargeN}) to create a data collapse. As it can be seen in the figure, collected data lay on the diagonal $z = \mu^ay^bN^c$.
\begin{figure}[!htb]
\begin{center}
\includegraphics[height=\columnwidth,angle=270]{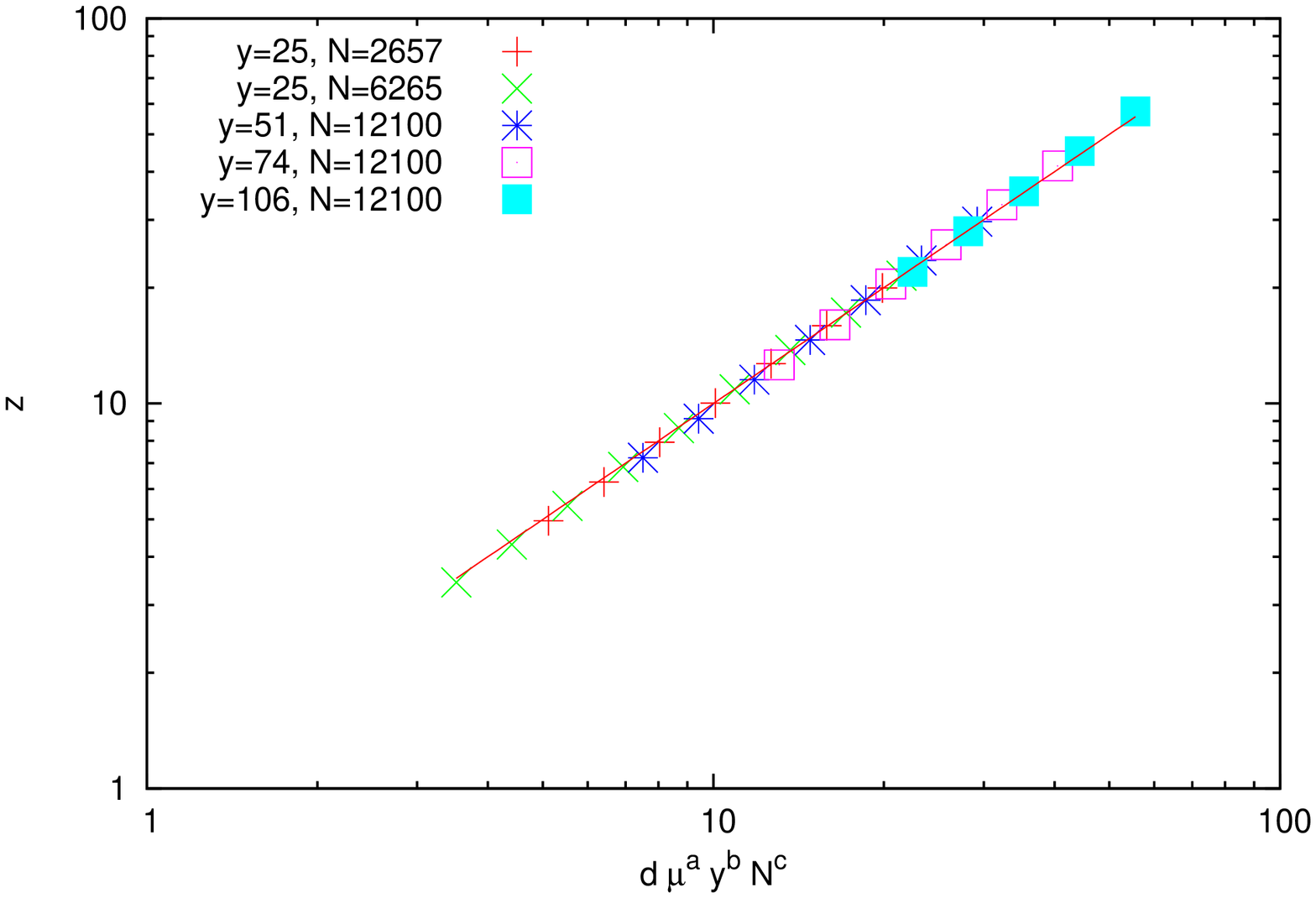}
\caption{Turnover rate $z$ vs $ d \cdot \mu^a y^b N^c$ for the Wright-Fisher model in the region $N \mu > y$, $y < N/100$. Values used were $a=0.558$, $b=0.879$, $c=0.091$ and $d=1.79$, the best fit values found \tref{WFreslargeN}. Error bars are smaller than symbols.}
\label{turnovercollapse_bigN}
\end{center}
\end{figure}
The changes in the coefficients of $a$ and $b$ for this large $N$ data of \tref{WFreslargeN} are statistically significant but not especially large. Using them to derive values for $\mu$ and $y$ from a real data set is not likely to cause difficulties. The problem lies with the dependence on $N$, as the power $c$ in $N^c$ varies significantly and so leads to large difference in numerical estimates from the fitted forms. These values for large $N$ in \tref{WFreslargeN} are again not compatible with the simple form \tref{newturnover} proposed in \cite{BLHH07}.  However the form $z = d y^a \mu^b N^c$ using the values derived from large $N$ \tref{WFreshighmuN}  now reproduce the data in table \ref{bigdata} for very large $N$ within 6\%. It should be noted, though, that in the $N \in [1000, 13000]$ region used for \tref{WFreslargeN} results the dependence of the turnover rate $z$ on the population size $N$ is weaker than in the first fit performed. This might suggest a logarithmic dependence of the turnover rate $z$ on the population size $N$ as for $c \ll 1$: $N^c \simeq 1+c\cdot \ln N$. Fits with this form work just as well as the one with the $N^c$ dependence.

\section{The Moran model}
In the Wright-Fisher model at each time step all individuals are assigned a new artefact simultaneously. While this can be a good update rule in some situations, it is certainly reasonable to think that cultural transmission and copying between individuals might occur in a more gradual way, with individuals changing taste individually one at a time.

This behaviour can be introduced in the model by changing the update rule so that at each time step only one randomly chosen individual is assigned a new artefact, by either copying it from an other individual in the population (chosen at random) or by inventing a new one with probability $\mu$. This model is known as ``Moran model'' \cite{E04}.

\begin{figure}
\begin{center}
\begin{tabular}{c|cccccc}
Individual	& 	&	t	&		& 	t+1	&	&	t+2\\ \hline
1	&	&	A	&		&	A	&	$\xrightarrow{\textit{innovate}}$	&	D\\
2	&	&	A	&		&	A	&	&	A\\
3	&	&	B	&	$\xrightarrow{\textit{copy 1}}$	&	A	&	&	A\\
4	&	&	A	&		&	A	&	&	A\\
5	&	&	C	&		&	C	&	&	C\\
6	&	&	A	&		&	A	&	&	A
\end{tabular}
\vspace{0.2cm}
\end{center}
\textbf{Top 3 chart}:\\
\begin{center}
\begin{tabular}{ccccccc}
Position	& 	&	t	&	& 	t+1	&	&	t+2	\\ \hline
1$^{\textit{st}}$		&	&	A	&		&	A	&	&	A\\
2$^{\textit{nd}}$	&	&	B	&	$\xrightarrow{\textit{ z = 1}}$	&	C	&	$\xrightarrow{\textit{ z = 1}}$	&	C	\\
3$^{\textit{rd}}$		&	&	C	&		&		&	&	D\\
\end{tabular}\\
\caption{A simple representation of the Wright-Fisher model. Shown are two successive time steps for a population of six individuals and the top chart at each time step.}
\end{center}
\label{moranrepr}
\end{figure}

In this section the regularity of turnover in the top $y$ chart of the most popular artefacts in the population for the Moran model is investigated as it has been done for the Wright-Fisher model in the previous section. The turnover $z$ is defined as before and most of the analysis is performed in the same way. Unless otherwise stated it is assumed that the investigation is conducted in the exact same way as for the Wright-Fisher model.

\subsection*{Steady state}

Analytical results \cite{E04,EP07} give that the typical time scale $\tau$ to reach a steady state is $\tau \sim  N \textrm{min}(\mu^{-1},N)$, differing from that in the Wright-Fisher model by a factor $N$ reflecting the different number of individuals it is possible to change each time step. This implies that simulations need to be run for a longer time with respect to the previous model. Again the steady state was confirmed numerically by studying $F_2(t)$ of \tref{F2def} which evolves in time in the same way as in the Wright-Fisher model \tref{F2evolve} but has a different eigenvalue controlling its evolution \cite{E04,EP07} $\lambda_2 = 1- 2 (\mu/N) - 2(1-\mu)/N^2$. For our parameter values we can safely choose $\Tsteady=4 N \mu^{-1}$ as the starting point to compute $\bar z$. Simulations have been run for $T=50+N\mu^{-1}$ time steps ($50$ time steps have been added to ensure a minimum amount of time steps for big values of $\mu$).

\subsection*{Data collection and analysis}

\begin{figure}[!hbt]
\begin{center}
\includegraphics[height=\columnwidth,angle=270]{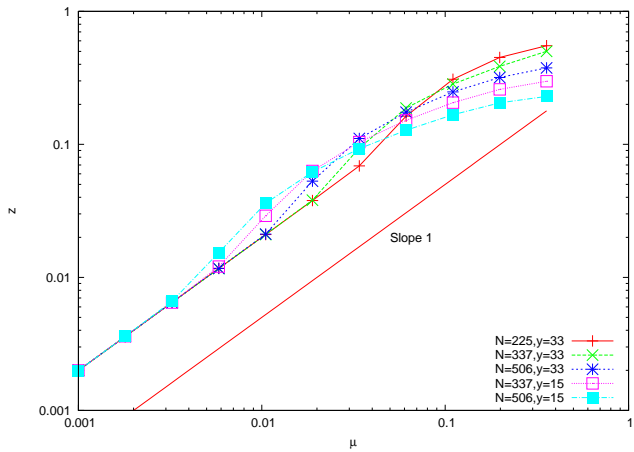}
\caption{Existence of a critical point. In the region $\mu < \mu_{c}$ we have $z \propto \mu$ ($y$ and $N$ are fixed for each of the plotted curves). Error bars are smaller than symbols.}
\label{moranz}
\end{center}
\end{figure}

\begin{figure}[!hbt]
\begin{center}
\includegraphics[height=\columnwidth,angle=270]{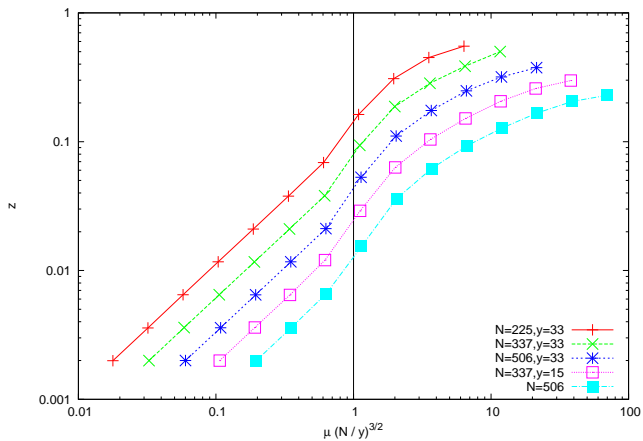}
\caption{Existence of a critical point $\mu_{c} \simeq (N/y)^{-3/2}$ ($y$ and $N$ are fixed for each of the plotted curves). Error bars are smaller than symbols.}
\label{moranzcritical}
\end{center}
\end{figure}

The usual \emph{ansatz} for the functional dependence of $z$ on $\mu$, $y$ and $N$ is made:
$z=d\cdot \mu^a y^b N^c$ \tref{funcform} and linear fits to $\ln(z)$ are used to estimate $a$, $b$, $c$ and $d$.
Approximately 350 data points have been collected in the following range $\mu \in [0.001, 0.357]$, $y \in [15,256]$ and $N \in [100, 506]$.  As figure \ref{moranz} shows, there is a transition point in $\mu$ that separates two different behaviours of the turnover rate $z$. This critical point is roughly $\mu_{c} \simeq \left(N/y \right)^{-1.5}$ as can be seen in figure \ref{moranzcritical}.
In the region $\mu < \mu_{c}$ the observed behaviour of the turnover rate $z$ is $z \propto \mu$, while for $\mu > \mu_{c}$ the turnover rate does not obey a simple power law as a function of the innovation rate $\mu$.

\subsubsection*{Region $\mu < \mu_c$}

Fitting the data points in this region with $z=d\cdot \mu^a y^b N^c$ \tref{funcform} we find
\beq
\begin{array}{cc}
a=0.9998(8) , &
b=-0.002(3) , \\
c=0.001(3)  , &
d=2.02(3)   ,
\end{array}
\label{Mreslowmun}
\eeq
so that $z=2 \cdot \mu$ within one standard deviation. In this region $z$ is independent on the top chart size $y$ and the population size $N$. This is best seen in figure \ref{moranz}.

\begin{figure}[!htb]
\begin{center}
\includegraphics[height=\columnwidth,angle=270]{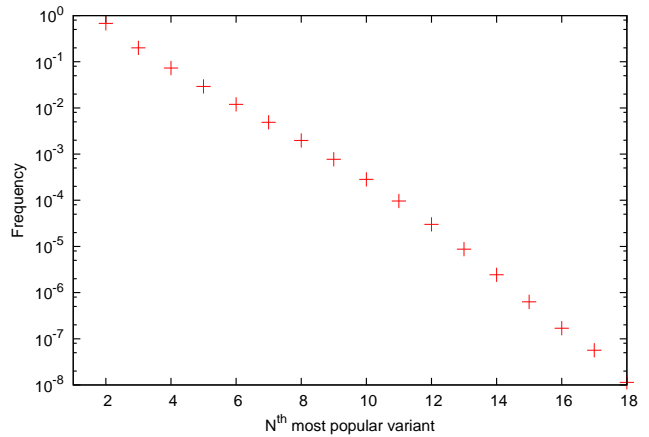}
\caption{Average frequency of artefacts (from the most popular one to the least popular). In the steady state there are no more than 18 artefacts in the population. With $N=268$, $\mu=0.00289$ and $y=140$ we have $\mu \ll \mu_c \simeq (N/y)^{-3/2}$ and the maximum number of artefacts in the population (18) is smaller than the top chart size $y$ (140).}
\label{moranfrequency}
\end{center}
\end{figure}

Keeping in mind that the average number of new artefacts introduced at each time step is equal to $\mu$ it is straightforward to understand that the mechanism that produces $z=2 \cdot \mu$ is similar to the one described for the Wright-Fisher model: for $\mu \ll \mu_c$ a very small number of artefacts survives in the population in the steady state. It is then likely that the number of artefacts in the population at a certain time step $t$ is lower than the top chart size $y$, as it is confirmed by observations (see figure \ref{moranfrequency}). Suppose now to let the population evolve for $N$ time steps, which is the unit of time in which all individuals in the population are assigned a new artefact. In this interval, on average, $N\cdot(1-\mu)$ individuals will copy another artefact in the population, most likely one of the most popular ones, while $N \mu$ individuals will invent a new artefact. At this point we will have a few very popular artefacts in the population and the $N \mu$ new artefacts recently introduced, all in the top chart. If we now take other $N$ time steps these very unpopular artefacts will likely be extinguished and replaced by other $N \mu$ invented artefacts. In the last $N$ time steps $N \mu$ artefacts exit the top chart and $N \mu$ new artefact enter it. The average turnover in these N time steps is therefore $z = \frac{N \mu + N \mu}{N} = 2\cdot \mu$

\subsubsection*{Region $\mu > \mu_c$}
\begin{figure}[!hbt]
\begin{center}
\includegraphics[height=\columnwidth,angle=270]{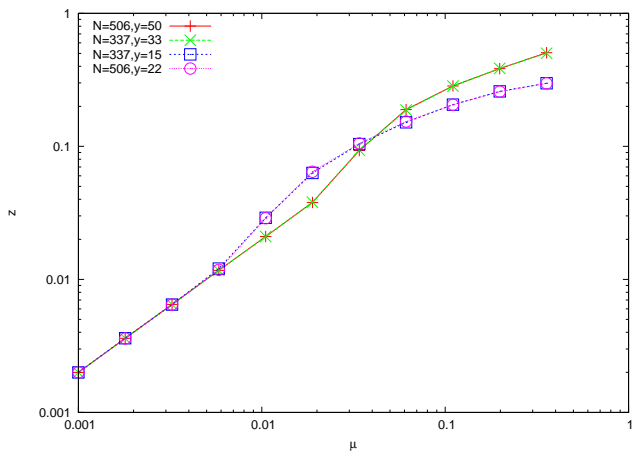}
\caption{$z$ is a function of y/N: curves with the same value of the ratio $y/N$ collapse. Error bars are smaller than symbols.}
\label{ysuN}
\end{center}
\end{figure}

We have seen in Fig.\ref{moranz}, above the critical point the hypothesis of a power law dependence of the turnover rate $z$ on the innovation rate $\mu$ is incorrect. Our results do suggest that the turnover rate $z$ is a function of the ratio $y/N$, as can be seen in figure \ref{ysuN}.

\section*{Conclusions}

In this paper we have investigated the dependence of the turnover rate $z$ on the parameters of the Wright-Fisher model and the Moran model.

We have found that in both models there is a critical point that separates two different regimes for the  dependence of $z$ on the innovation rate $\mu$, the top chart size $y$ and the population size $N$.
This critical point satisfies the equation $N \mu \simeq 0.15 \cdot y$ for the  Wright-Fisher model and $\mu \simeq (N/y)^{-3/2}$ for the Moran model.  The two models behave similarly below the critical point, where the turnover rate is given by two times the average number of new artefacts entering the population in one time step. This is equal to $N\mu$ in the Wright-Fisher model and to $\mu$ in the Moran model.

Most data sets though will be above the critical region where we have seen that the two models behave differently, with the \textit{ansatz} $z=d\cdot \mu^a y^b N^c$ being a good approximation for the functional dependence of the turnover rate on the model variables for the Wright-Fisher model only.  The simple form $z=2\cdot \sqrt \mu \cdot y$ suggested by Bentley et al.\ \cite{BLHH07} is excluded statistically by our results.  We find that the powers of $\mu$ and $y$ differ by 10\% from the values in \cite{BLHH07} and that there is significant dependence on the size of the system $N$. In particular we have also shown that our fit in the region where the top chart size is much smaller than the population size ($y \ll N$) \tref{WFreslargeN} reproduces our simulation results for large populations within 6\%, and we believe that this should be used when extracting information from real data sets. One outstanding issue is that our two fits \tref{WFreshighmuN} and \tref{WFreslargeN} are not compatible in terms of the dependence of the turnover rate $z$ on $N$.  Further work is needed on this as this suggests the dependence of $z$ on $N$ may not be a simple power law.

For the Moran model for full top $y$ charts, the usual situation in real data, we have seen that the turnover doesn't follow a simple power law in $\mu$, $y$ and $N$. We have however noted that the turnover rate in the Moran model appears to be a function of the ratio $y/N$.

Since data on cultural transmission is sometimes available only in the form of popularity charts, the study of the turnover is of real practical use.  However the models considered here and in \cite{BLHH07} are only the simplest examples. It would be interesting to study the behaviour of the turnover in more complicated models such as examples which interpolate between the Wright-Fisher and Moran models \cite{EPY10,BOB11}. Another interesting avenue is to understand how the social network between individuals \cite{EPY10} alters the turnover.

\section*{Acknowledgements}

We thank the High Performance Computing Centre at Imperial College London for use of their cluster in the large $N$ studies.

\section*{Appendix}

The information in this appendix is supplementary material which will not appear in the published version.

\subsection*{Wright-Fisher model, small $N$}

\begin{figure}[!hbt]
\begin{center}
\includegraphics[height=\columnwidth,angle=270]{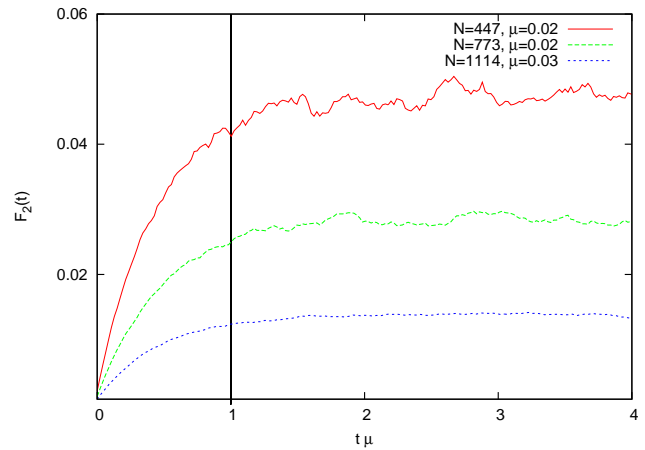}
\caption{Ensemble average of $\frac{\langle k(k-1) \rangle}{N(N-1)}$. The reach of a steady state can clearly be seen at $t \simeq 2 \mu^{-1}$.}
\label{figsteady}
\end{center}
\end{figure}

For the Wright-Fisher model, we used $F_2(t)$ to check our numerical simulations reach a steady state on the time scale as can be seen in Fig.\ \ref{figsteady}.

The observed approximate values of $\tau$ are listed in table \ref{tau} and compared with $\mu^{-1}$ in the two regions $N \mu < 1$ and $N \mu \geq 1$: it is found that the relation $\tau \sim \mu^{-1}$ holds in the region $N \mu \geq 1$, while $\tau < \mu^{-1}$ for $N \mu < 1$. The observed behaviour of $\tau$ in the second region can be due to the fact that being less than one innovator per generation ($N \mu < 1$) on average, there is less variability in the artefacts and therefore a fastest approach to steady state through copying.
\begin{table}[htdp]
\begin{tabular}{c|ccccc|c}\backslashbox{$\mu$}{$N$} & $50$ & $100$ & $500$ & $1000$ & $1500$ & $\mu^{-1}$\\ \hline
$0.02$ & $ 50$ & $50$ & $50$ & $50$ & $50$ & $50$\\
$0.01$ & $ 80$ & $100$ & $100$ & $100$ & $100$ & $100$ \\
$0.002$ & $ 250$ & $ 300$ & $500$ & $500$ & $500$ & $500$\\
$0.001$ & $ 300$ & $ 400$ & $ 800$ & $1000$ & $1000$ & $1000$\\
$0.00067$ & $ 300$ & $ 300$ & $ 1000$ & $ 1500$ & $1500$ & $1500$
\end{tabular}
\caption{Approximate values of $\tau$. Above the main diagonal $N \mu \geq 1$, below  $N \mu < 1$.}
\label{tau}
\end{table}

The data was fitted to \tref{funcform} by applying the linear fit routine $fit$ in \textit{Gnuplot} to the logarithm of $z$ and the variables $\mu$, $y$ and $N$
\begin{equation}
\ \ln z = a \cdot \ln \mu + b \cdot \ln y + c \cdot \ln N + \ln d \, .
\label{linearfit}
\end{equation}
on collected data points. For this purpose code and scripts have been set up to record $\ln \langle \bar z \rangle$ and $\sigma_{\ln \langle \bar z \rangle }$.

To perform the fit with equally spaced data $\mu$, $y$ and $N$ are varied in the following way $\mu = \mu_0 \cdot q^m$, $y = y_0 \cdot r^n$ and $N = N_0 \cdot s^p$ with fixed $\mu_0$, $y_0$, $N_0$, $q$, $r$, $s$ and $m,n,p = 0, 1, 2\dots$. We choose $q=1.4$, $r=1.3$, $s=1.2$.

\subsection*{Moran  model}

\begin{figure}[!hbt]
\begin{center}
\includegraphics[height=\columnwidth,angle=270]{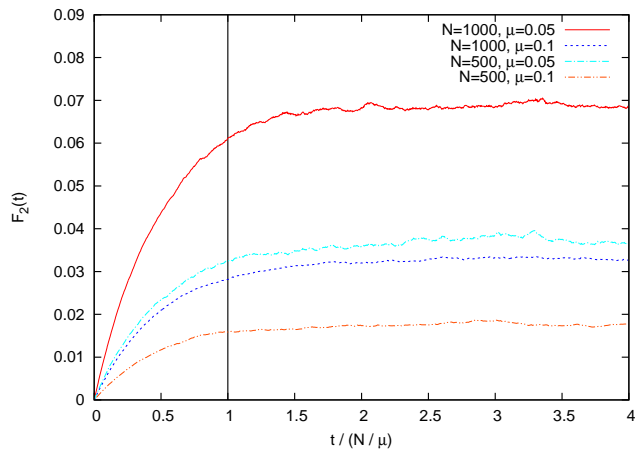}
\caption{Ensemble average of $\frac{\langle k(k-1) \rangle}{N(N-1)}$. The reach of a steady state can clearly be seen at $t \simeq 2 N \mu^{-1}$.}
\label{figsteadymoran}
\end{center}
\end{figure}

We have from analytic results \cite{E04,EP07} that the time scale for equilibrium is $\tau^{-1} \sim \ln(\lambda_2) \lesssim N \textrm{min}((2\mu)^{-1},N)$.  This can be seen in our numerical simulations in figure \ref{figsteadymoran}.

To perform the fit with equally spaced data $\mu$, $y$ and $N$ are varied in the following way:
$\mu = \mu_0 \cdot q^m$, $y = y_0 \cdot r^n$ and $N = N_0 \cdot s^p$
with fixed $\mu_0$, $y_0$, $N_0$, $q=1.5$, $r=1.8$, $s=1.5$ and $n= 0, 1, 2\dots$. A further constraint is that the top chart size satisfies $y<N$.


\end{document}